\documentstyle[prl,aps,epsfig,amssymb]{revtex}

\begin{document}
\newcommand{\beq}{\begin{equation}}
\newcommand{\eeq}{\end{equation}}
\newcommand{\barr}{\begin{eqnarray}}
\newcommand{\earr}{\end{eqnarray}}

\newcommand{\andy}[1]{ }

\def\rz{\mbox{\boldmath $y$}_0}
\def\a{\alpha_0} \def\da{\delta\alpha}
\def\Da{D_\alpha}
\def\h{\widehat}
\def\t{\widetilde}
\def\cH{{\cal H}}
\def\ud{\uparrow\downarrow}
\def\uu{\uparrow}
\def\dd{\downarrow}
\def\u{\uparrow}
\def\d{\downarrow}
\def\b1{{\bf 1}}
\def\chiA{\chi_{_{A}}}
\newcommand{\bm}[1]{\mbox{\boldmath $#1$}}
\newcommand{\bmsub}[1]{\mbox{\boldmath\scriptsize $#1$}}
\newcommand{\bmh}[1]{\mbox{\boldmath $\hat{#1}$}}

\def\coltwovector#1#2{\left({#1\atop#2}\right)}
\def\upp{\coltwovector10}    \def\downn{\coltwovector01}
\def\bra#1{\langle #1 |}
\def\ket#1{| #1 \rangle}
%
%
\def\ask{\marginpar{?? ask:  \hfill}}
\def\fin{\marginpar{fill in ... \hfill}}
\def\spnote{\marginpar{note (SP) \hfill}}
\def\pfnote{\marginpar{note (Paolo) \hfill}}
\def\check{\marginpar{check \hfill}}
\def\discuss{\marginpar{discuss \hfill}}
%
\def\lsnote#1{$\bullet\bigl\{$ {\def\dash{\hbox{\rm---}}\tt#1} $\bigr\}
\bullet$
\marginpar{note~(LS) \hfill}}
\def\NI{\noindent }
\def\ellip{ $\ldots$ }
\def\eqcite#1{(\ref{eq:#1})}
\def\headerls#1{\bigskip\noindent{\bf #1}\newline\smallskip}
 \def\Tr{\mathop{\rm Tr}\nolimits}
\font\romeight=cmr8 scaled\magstep0  \font\boldeight=cmbx8
scaled\magstep0
\font\italeight=cmti8 scaled\magstep0 \font\sleight=cmsl8 scaled
\magstep0
\font\tteight=cmtt8 scaled \magstep0
\def\smallletters{\romeight\boldeight\italeight\sleight\def\tt{\tteight}
\def\bf{\boldeight}\def\it{\italeight}\def\rm{\romeight}}


\draft

\title{ Zeno dynamics yields ordinary constraints}
\author{ P. Facchi,$^{1}$ S. Pascazio,$^{2,3}$ A. Scardicchio$^{2}$ and
L. S. Schulman$^{4}$}

\address{$^{1}$Atominstitut der \"Osterreichischen Universit\"aten,
Stadionallee
2, A-1020, Wien, Austria \\
 $^{2}$Dipartimento di Fisica, Universit\`a di Bari  I-70126 Bari, Italy
\\
$^{3}$Istituto Nazionale di Fisica Nucleare, Sezione di Bari,  I-70126
Bari,
Italy \\
$^{4}$Physics Department, Clarkson University, Potsdam, NY 13699-5820,
USA }

\date{\today}

\maketitle

\begin{abstract}
The dynamics of a quantum system undergoing frequent measurements
(quantum Zeno effect) is  investigated. Using asymptotic analysis, the
system is found to evolve unitarily in a proper subspace of the total
Hilbert space. For spatial projections, the generator of the ``Zeno
dynamics" is the Hamiltonian with Dirichlet boundary conditions.
\end{abstract}


Frequent measurement can slow the time evolution of a quantum system,
hindering transitions to states different from the initial one
\cite{Beskow,Misra}. This phenomenon, known as the quantum Zeno effect
(QZE), follows from general features of the Schr\"odinger equation that
yield quadratic behavior of the survival probability at short times
\cite{strev}.

However, a series of measurements does not necessarily freeze
everything. On the contrary, for a projection onto a multi-dimensional
subspace, the system can evolve away from its initial state, although it
remains in the subspace defined by the ``measurement." This continuing
time evolution {\em within} the projected subspace we call {\em quantum
Zeno dynamics}. It is often overlooked, although it is readily
understandable in terms of a theorem on the QZE \cite{Misra} that we
will recall below.

The aim of this article is to show that Zeno dynamics yields
ordinary constraints. Under general conditions, the evolution of
a system undergoing frequent measurements takes place in a proper
subspace of the total Hilbert space and the wave function
satisfies Dirichlet boundary conditions on the domain defined by
the measurement process. Moreover, the evolution is generated by
a self-adjoint Hamiltonian and remains reversible within the Zeno
subspace. This shows that the irreversibility is not compulsory,
as noted in \cite{FGMPS}.

The QZE has been tested on oscillating systems \cite{Cook} and there has
been a recent observation of non-exponential decay (leakage through a
confining potential) at short times \cite{Wilkinson}. Although these
experiments have invigorated studies on this issue, they deal with
one-dimensional projectors (and therefore one-dimensional Zeno
subspaces): the system is forced to remain in its initial state. This is
also true for interesting quantum optical applications \cite{perina}.
The present work therefore enters an experimentally uncharted area,
although the property of being a multidimensional measurement is not at
all exotic, and in particular applies to the most basic quantum
measurement: position. The latter is the paradigm for the present work.

We introduce notation. Consider a quantum system, Q, whose states belong
to the Hilbert space ${\cal H}$ and whose evolution is described by the
unitary operator $U(t)=\exp(-iHt)$, where $H$ is a time-independent
semi-bounded Hamiltonian. Let $E$ be a projection operator that does not
commute with the Hamiltonian, $[E,H]\neq 0$, and $E{\cal H}={\cal H}_E$
the subspace defined by it. The initial density matrix $\rho_0$ of
system Q is taken to belong to ${\cal H}_E$: \andy{inprep}
\beq
\rho_0 = E \rho_0 E , \qquad \Tr \rho_0 = 1 \,.
\label{eq:inprep}
\eeq
The state of Q after a series of $E$-observations at times $t_j=jT/N \;
(j=1,
\cdots, N)$ is \andy{Nproie}
\beq \rho^{(N)}(T) = V_N(T) \rho_0 V_N^\dagger(T) , \qquad
    V_N(T) \equiv [ E U(T/N) E ]^N
\label{eq:Nproie}
\eeq
and the probability to find the system in ${\cal H}_E$ (``survival
probability")
is \andy{probNob}
\beq
P^{(N)}(T) = \mbox{Tr} \left[ V_N(T) \rho_0 V_N^\dagger(T) \right].
\label{eq:probNob}
\eeq
Our attention is focused on the limiting operator \andy{slim}
\beq {\cal V} (T) \equiv \lim_{N
\rightarrow \infty} V_N(T).
  \label{eq:slim}
\eeq
Misra and Sudarshan \cite{Misra} proved that if the limit exists, then
the
operators ${\cal V} (T)$ form a one-parameter semigroup, and the final
state
is \andy{infproie}
\beq
\rho (T)= \lim_{N \rightarrow \infty} \rho_N(T)
= {\cal V}(T) \rho_0 {\cal V}^\dagger (T) .
  \label{eq:infproie}
\eeq
The probability to find the system in ${\cal H}_E$ is \andy{probinfob}
\beq
{\cal P} (T) \equiv \lim_{N \rightarrow \infty} P^{(N)}(T)
   = 1.
\label{eq:probinfob} \eeq
This is the QZE. If the particle is constantly checked for whether it
has
remained in ${\cal H}_E$, it never makes a transition to $({\cal
H}_E)^\perp$.

A few comments are in order. First, the final state $\rho (T)$ depends
on the characteristics of the model investigated and on the measurement
performed (the specific forms of $V_N$ and ${\cal V}$ depend on $E$).
Moreover, the physical mechanism that ensures the conservation of
probabilities within the relevant subspace hinges on the short time
behavior of the survival probability: probability leaks out of the
subspace ${\cal H}_E$ like $t^2$ for short times. Since the infinite-$N$
limit suppresses this loss, one can inquire under what circumstances
${\cal V}(T)$ actually forms a group, yielding reversible dynamics
within the Zeno subspace.

In this article we show that Zeno dynamics for a position
measurement yields a particular kind of dynamics within the subspace
defined by that measurement, namely unitary evolution with the
restricted Hamiltonian, and with the domain of that (self adjoint)
operator defined by Dirichlet boundary conditions. This elucidates the
reversible features of the evolution for a wide class of physical
models. As a spinoff, our proof provides a rigorous regularization of
the example considered in \cite{FGMPS}.

We start with the simplest spatial projection. Q is a free particle of
mass $m$ on the real line, and the measurement is a determination of
whether or not it is in the interval $A=[0,L]\subset \mathbb{R}$. The
Hamiltonian and the corresponding evolution operator are \andy{freeHam}
\beq H=\frac{p^2}{2m}\,,\qquad U(t)=\exp(-i t H )\,.
\label{eq:freeHam}
\eeq
$H$ is a positive-definite self-adjoint operator on $L^2(\mathbb{R})$
and $U(t)$ is unitary. We study the evolution of the particle when it
undergoes frequent measurements defined by the projector \andy{proja}
\beq
E_{A}=\int dx\;\chiA(x) \ket{x}\bra{x},
\label{eq:proja}
\eeq
where $\chiA$ is the characteristic function \andy{chara}
\beq
\chiA(x)=\left\{
  \begin{array}{l}
    1 \quad \mbox{for } x \in A =[0,L] \\
    0 \quad \mbox{otherwise}
  \end{array}\right. .
\label{eq:chara}
\eeq
Thus $E_A$ is the multiplication operator by the function $\chi_A$. We
study the following process. We prepare a particle in a state with
support in $A$, let it evolve under the action of its Hamiltonian,
perform frequent $E_A$ measurements during the time interval $[0,T]$,
and study the evolution of the system {\em within} the subspace ${\cal
H}_{E_A}=E_A{\cal H}$. We will show that the dynamics in ${\cal
H}_{E_A}$ is governed by the evolution operator
\andy{VT,HZ}
\barr
& & {\cal V}(T)=\exp(-iT H_{\rm Z})\, E_A,
\label{eq:VT} \\
& & \qquad\mbox{with}\qquad H_{\rm Z}\equiv\frac{p^2}{2m}+V_A(x),
\quad V_A(x)=\left\{
  \begin{array}{l}
      0 \qquad \mbox{~for } x \in A \\
      +\infty \quad \mbox{otherwise}
  \end{array}\right. .
\label{eq:HZ}
\earr
This is the operator obtained in the limit (\ref{eq:slim}). In other
words, the system behaves as if it were confined in $A$ by rigid walls,
inducing the wave function to vanish on the boundaries $x=0,L$
(Dirichlet boundary conditions).

We now prove our assertion. Let the particle be initially ($t=0$) in
$A$. We recall the propagator in the position representation
\cite{FeynmanHibbs,Schulman} \andy{euclisingle}
\barr
G(x,t;y) & \equiv & \bra{x} E_A\, U(t) E_A
\ket{y}=\chiA(x)\bra{x}U(t)\ket{y}\chiA(y)
\nonumber\\
&=& \chiA(x) \sqrt{\frac{m}{2\pi it
\hbar}}\exp\left[\frac{im(x-y)^2}{2\hbar t}\right]\chiA(y),
\label{eq:euclisingle}
\earr
where $t=T/N$ is the time when the first measurement is carried
out and the particle found in $E_A$. To study the properties of
$G$ we choose a complete basis in $L^2(A)$ \andy{basis}
\beq
u_n (x)= \langle x|u_n\rangle = \sqrt{\frac{2}{L}}
\sin\left(\frac{n\pi x}{L}\right) \quad (n=1,2,\ldots) \;.
\label{eq:basis}
\eeq
When these functions define the eigenbasis of $H$, $H$ is self
adjoint and
\beq
H|u_n\rangle = E_n |u_n\rangle \;, \quad
E_n=\frac{\hbar^2n^2\pi^2}{2mL^2}\;,
\label{eq:base}
\eeq
so that $H$ has Dirichlet boundary conditions. The matrix elements of
$G$ are
\andy{Gmn}
\barr
G_{mn}(t)  \equiv
\langle u_m |E_A U(t)E_A|u_n\rangle  =
\int_0^L dx  \int_0^Ldy \; u_m(x) \,
 \sqrt{\frac{m}{2\pi it \hbar}}\exp\left[\frac{im(x-y)^2}{2\hbar
t}\right]
\,  u_n(y)       \;.
\label{eq:Gmn}
\earr
Let $r=x-y$, $R=(x+y)/2$ and $\lambda=m/2\hbar t$, so that
\andy{Gmnr}
\beq
G_{mn}(\lambda)=\sqrt{\frac{\lambda}{i\pi}}\int_0^LdR
\int_{-r_0(R)}^{r_0(R)}dr \; u_m(R+r/2)u_n(R-r/2) \exp\left[i\lambda
r^2\right].
\label{eq:Gmnr}
\eeq
where $r_0(R)=L-|L-2R|$. We now use the asymptotic expansion
\andy{asy}
\beq
g(\lambda)=\sqrt{\frac{\lambda}{i\pi}}\int_{-a}^a dx \; f(x) e^{i\lambda
x^2} =
g_{\rm stat}(\lambda) + g_{\rm bound}(\lambda),
\label{eq:asy}
\eeq
where
\andy{asyborder1,2}
\barr
g_{\rm stat}(\lambda) & = & f(0) + \frac{i}{4\lambda} f''(0) +
O(\lambda^{-2}), \label{eq:asyborder1}
\\ g_{\rm bound}(\lambda) & = & \frac{e^{i\lambda
a^2}}{2ia \sqrt{i\pi\lambda}}[f(a)+f(-a)] + O(\lambda^{-3/2})
\label{eq:asyborder2}
\earr
are the contributions of the stationary point $x=0$ and of the boundary,
respectively. By expanding the inner integral in (\ref{eq:Gmnr}) as in
(\ref{eq:asy})--(\ref{eq:asyborder2}) one gets \andy{Gmnrlam}
\barr
& & \sqrt{\frac{\lambda}{i\pi}}
\int_{-r_0(R)}^{r_0(R)}dr \; u_m(R+r/2)u_n(R-r/2) \exp\left[i\lambda
r^2\right]
\nonumber \\ & &
= u_m(R)u_n(R) +\frac{i}{4 \lambda}
\frac{d^2}{dr^2} \left[ u_m(R+r/2)u_n(R-r/2) \right]_{r=0}+
O(\lambda^{-3/2})
\,.
\label{eq:Gmnrlam}
\earr
(Note that the contribution of the boundary  vanishes identically.)
Using this result, we integrate by parts and after a straightforward
calculation obtain
\andy{Gmnlamb}
\barr
G_{mn}(t)&=&\int_0^L dR \left[u_m(R)u_n(R) -\frac{it}{\hbar}
u_m(R)
\frac{-\hbar^2}{2m} \frac{d^2}{dR^2} u_n(R) \right]+ O(t^{3/2})
\nonumber \\
& = & \langle u_m|u_n \rangle - \frac{it}{\hbar} \langle
u_m|\frac{p^2}{2m}|u_n \rangle + O(t^{3/2}) \nonumber \\
& = & \delta_{mn} \left( 1 -\frac{it}{\hbar} E_n \right) +
O(t^{3/2})\,.
\label{eq:Gmnlamb}
\earr
With this formula we can carry out the limit required in
Eq.~(\ref{eq:slim}). At time $T$, in the energy representation, the
propagator becomes \andy{GmnT}
\barr
{\cal G}_{mn}(T)&=&\bra{m} {\cal V}(T)\ket{n} \nonumber \\
& = & \lim_{N \to \infty} \sum_{n_1\ldots n_{N-1}}
G_{mn_1}(T/N)G_{n_1n_2}(T/N) \cdots G_{n_{N-1}n}(T/N)
\nonumber \\
& = & \delta_{mn} e^{-iTE_n/\hbar}.
\label{eq:GmnT}
\earr
This is precisely the propagator of a particle in a square well with
Dirichlet boundary conditions and proves (\ref{eq:VT})--(\ref{eq:HZ}).
Note that the $t^{3/2}$ contribution in (\ref{eq:Gmnlamb}) drops out in
the $N\to\infty$ limit since it appears as $N \times O(1/N^{3/2})$. It
is worth emphasizing that although this result has been proved using the
basis (\ref{eq:basis}), the information obtained is a property of the
propagator, and therefore holds true in general. Our choice was a matter
of convenience. With a different basis and nonvanishing boundary
conditions, the dominant contribution of order $\lambda^{-1/2}$ in
$g_{\rm bound}(\lambda)$ would have given a nondiagonal term in
(\ref{eq:Gmnrlam})--(\ref{eq:GmnT}), showing that the chosen basis is
not the right eigenbasis of $H_{\rm Z}$ (i.e., for the limiting object).

This result can be generalized to a wide class of systems. Let
\andy{HamV}
\beq
H=\frac{p^2}{2m}+V ,\qquad U(t)=\exp(-i t H ),
\label{eq:HamV}
\eeq
where $V$ is a regular potential. (It may be unbounded from below, for
example $V(x)=Fx$, although within $A$ the total Hamiltonian $H$ should
be lower bounded.) The measurement performed is again application of the
projector (\ref{eq:proja}) and we study the short-time propagator
\andy{euclisingleV}
\barr
G(x,t;y)  &=& \chiA(x) \sqrt{\frac{m}{2\pi it
\hbar}}\exp\left[\frac{im(x-y)^2}{2\hbar t}\right]
\exp\left[-\frac{it(V(x)+V(y))}{2\hbar}\right] \chiA(y).
\label{eq:euclisingleV}
\earr
The basis to be used for representing the propagator is again
that of the Hamiltonian with Dirichlet boundary conditions in
$[0,L]$
\andy{baseV}
\beq
H|u_n\rangle = \left(\frac{p^2}{2m}+V \right) |u_n\rangle =  E_n
|u_n\rangle\, ,
 \quad u_n(x)|_{x=0,L}=0 .
\label{eq:baseV}
\eeq
As before ($r=x-y, \; R=(x+y)/2, \; \lambda=m/2\hbar t$)
\andy{GmnrV}
\beq
G_{mn}(\lambda)=\sqrt{\frac{\lambda}{i\pi}}\int_0^L dR
\int_{-r_0(R)}^{r_0(R)}dr \; u_n\left(R+\frac r2\right)
e^{-itV(R+ r/2)/2\hbar} u_n\left(R-\frac r2\right) e^{-itV(R-
r/2)/2\hbar}
e^{i\lambda r^2} .
\label{eq:GmnrV}
\eeq
Using the asymptotic expansion (\ref{eq:asy})--(\ref{eq:asyborder2}), a
calculation identical to the previous one yields \andy{GmnlambV}
\barr
G_{mn}(t)&=&\int_0^L dR \left[u_n(R)u_m(R) -\frac{it}{\hbar}
u_n(R)
\left(\frac{-\hbar^2}{2m} \frac{d^2}{dR^2}+ V(R) \right)
u_m(R) \right]+ O(t^{3/2}) \nonumber \\
& = & \langle u_n|u_m \rangle - \frac{it}{\hbar} \langle
u_n|\left(\frac{p^2}{2m} + V \right)|u_m \rangle + O(t^{3/2})
\nonumber
\\
& = & \delta_{nm} \left( 1 -\frac{it}{\hbar} E_n \right) +
O(t^{3/2})
\label{eq:GmnlambV}
\earr
and the limiting propagator at time $T$ again reads
\andy{GmnTV}
\beq
{\cal G}_{mn}(T)=\delta_{nm} e^{-iTE_n/\hbar}.
\label{eq:GmnTV}
\eeq
Again, the simplicity of the proof is due to the choice of the
basis (\ref{eq:baseV}), satisfying Dirichlet boundary conditions.

We have also obtained an improvement with respect to earlier approaches
to this problem. The aforementioned theorem by Misra and Sudarshan
\cite{Misra} requires that the Hamiltonian be lower bounded from the
outset. However, we need only require that the Hamiltonian be lower
bounded in the Zeno subspace. Despite the fact that for unbounded
potentials (like $V=Fx$) $H$ may not be lower bounded on the real line,
the evolution in the Zeno subspace is governed by the Hamiltonian
\andy{HZV}
\beq
H_{\rm Z}=\frac{p^2}{2m}+V_A(x),
\quad V_A(x)=\left\{
  \begin{array}{l}
    V(x) \quad \mbox{for } x \in A \\
    +\infty \quad \mbox{otherwise}
  \end{array}\right.
\label{eq:HZV}
\eeq
that can be lower bounded in $A$, yielding a {\em bona fide} group for
the evolution operators.

The above calculation and conclusions can readily be generalized to
higher dimensions, so long as the measurement projects onto a set in
${\mathbb R}^n$ with a smooth boundary (except at most a finite number
of points). We again take $\bm x, \bm y \in {\mathbb R}^n$ and let the
measurement-projection be defined by $A\subset {\mathbb R}^n$, which is
not necessarily bounded. Again setting $\bm r=\bm x-\bm y, \; \bm R=(\bm
x+\bm y)/2$, Eq.\ (\ref{eq:GmnrV}) becomes
\andy{GmnrVn}
\beq
G_{mn}(\lambda)=\left(\frac{\lambda}{i\pi}\right)^{n/2}\int_A d
\bm R
\int_{D(\bmsub R)}d \bm r \; u_n(\bm R+ \bm r/2)
e^{-itV(\bmsub R+\bmsub r/2)/2\hbar} u_n(\bm R-\bm r/2)
e^{-itV(\bmsub R-\bmsub r/2)/2\hbar} e^{i\lambda \bmsub r^2} ,
\label{eq:GmnrVn}
\eeq
where $D(\bm R)$ is the transformed integration domain for $\bm r$. The
$n$-dimensional asymptotic expansions
(\ref{eq:asy})--(\ref{eq:asyborder2}) read \cite{Bleistein}
\andy{asybordern1,2}
\barr
g_{\rm stat}(\lambda) & = & f(\bm 0) + \frac{i}{4 \lambda}
\; \triangle f(\bm 0) + O(\lambda^{-2}),
\label{eq:asybordern1}
\\ g_{\rm bound}(\lambda) & = & O(\lambda^{-1/2}) \times f({\rm
boundary})+
O(\lambda^{-3/2})
 \label{eq:asybordern2}
\earr
and the theorem follows again because $f$ vanishes on the boundary
(Dirichlet). The proof is readily generalized to non-convex and/or
multiply-connected projection domains, the only difficulty being
that the integration domain in (\ref{eq:GmnrVn}) must be broken
up. It is interesting to notice that at those points at which the
boundary fails to have a continuously turning tangent plane, the
asymptotic contribution of the discontinuity in the boundary in
(\ref{eq:asybordern2}) would be of yet higher order in $\lambda$.

In conclusion, for traditional position measurements, namely
projections onto spatial regions, we have shown that Zeno
dynamics uniquely determines the boundary conditions, and that
they turn out to be of Dirichlet type. This is also relevant for
problems related to the consistent histories approach
\cite{Yamada,Hartle,Muga}, where different boundary conditions
were proposed. For us, the frequent imposition of a projection,
the traditional idealization of a measurement, provides all the
decohering of interfering alternatives that is needed. On the
other hand, in the works just cited one seeks a restricted
propagator (using the path decomposition expansion
\cite{auerbach}) and such interference can occur. A second issue
discussed in these works (especially \cite{Hartle}) is the
validity of the Trotter product formula in certain cases. Again,
our implicit use of this formula (in Eq.~(\ref{eq:euclisingle}),
etc.) is nothing more than its use for a free particle (or a
particle in an ordinary potential). This is because the
propagator of Eq.~(\ref{eq:euclisingle}) provides time evolution
under a {\em sequence} of operations: the particle evolves freely
(on the entire line) for a time $t$, and {\em then} one applies
the projection (left and right multiplication by the operator
$E_A$).

The present work has implications for the notion of ``hard wall," as
used for example in elementary quantum mechanics.  Everyone would agree
(we expect) that this notion is an idealization. However, in many cases
where this idealization is useful the ``wall" is dynamic rather than
static, the result of some fluctuating atomic presence. In this article
we have a sufficient condition for the validity of this notion in a
dynamic situation. Moreover, there is a quantitative framework (arising
from our asymptotic analysis and finite-time-interval QZE effects) for
gauging the effects of less than perfect hard walls.


\section*{Acknowledgments}
This work is supported in part by the TMR-Network of the European Union
``Perfect Crystal Neutron Optics" ERB-FMRX-CT96-0057 and by the United
States NSF under grant PHY 97 21459.


\end{document}